\input{aipcheck}

\documentclass[
    ,final            
  ]
  {aipproc}

\layoutstyle{6x9}

\begin{document}

\title{RECENT PROGRESS ON COLOR CONFINEMENT}
\author{Adriano Di Giacomo}{
 address={Physics Dept. Pisa University and I.N.F.N.Sezione di Pisa}}
\begin{abstract}
The status is reviewed of the dual superconductivity of QCD vacuum as a mechanism of color confinement.
 \end{abstract}
\maketitle


\section{INTRODUCTION}

   The existing experimental evidence for confinement of quarks is striking.
   
    The ratio of the abundance of quarks to that of nucleons in nature   $a_q\equiv {n_q\over n_p}$    has an experimental upper limit    $a_q\le 10^{-27}$    \cite{PDG} , to be compared to the prediction of the Standard Cosmological model   $a_q\approx 10^{-12}$  in the hypothesis of no confinement \cite{Okun}. 
    
    Similarly the cross section for inclusive production of quarks + antiquarks     $\sigma_q\equiv  \sigma(p+p\to q(\bar{q}) + X)$ has an experimental upper limit  $ \sigma_q\le 10^{-40}cm^2 $ \cite{PDG}, to be compared to the prediction of perturbation theory for unconfined quarks  $\sigma_q\approx \sigma_{TOT} \approx 10^{-25} cm^2$.
    
    The inhibition factor is  $\le 10^{-15}$  which is a very small number. The natural explanation would then be that  quarks are absolutely confined, i.e. that these quantities are strictly zero, due to some symmetry which is responsible for confinement. This would imply in turn that the deconfining transition is a change of symmetry  , i.e. an order-disorder transition : it cannot be a crossover. A continuous crossover would mean that one can go continuously from confined to deconfined  phase, and would therefore require an (unnatural ) explanation of the factor $10^{-15}$.  Of course the issue is finally  decided by Nature, but this is a fundamental point to be investigated.  In pure gauge theories , (no quarks), lattice calculations show that deconfinement is indeed an order disorder transition and an order parameter is the Polyakov line .  In the presence of quarks the question is more controversial. A brief discussion of the status of this problem, which is fundamental, will be   part of this talk.   
    
    I will also discuss one candidate  symmetry to explain confinement :  dual superconductivity  of the vacuum \cite{'tHooft1}. The idea is that vacuum is a dual superconductor below the deconfining temperature $T_c$ , and becomes normal above $T_c$ .  In the confined phase magnetic charges condense so that vacuum has no definite magnetic charge but is a superposition of states with different magnetic charges. In the deconfined phase, instead, magnetic charge is superselected.
    Confinement is produced via (dual) Meissner effect : the electric field acting between a $q-\bar q$ pair
    is channeled into Abrikosov flux tubes and the energy is proportional to the distance.
    
    Traditionally there are two distinct strategies to investigate this phenomenon.
    
     1) Look at the symmetry.  This amounts to define an order parameter  $<\mu>$ , which is the expectation value of a magnetically charged operator  $\mu$ .If the mechanism of dual superconductivity  is at work one  expects  $<\mu> \neq 0$ in the broken phase, $<\mu>  = 0$  in the deconfined phase \cite{DigZak} \cite{DelDigPafPier}\cite{Frolich}.
     
     2) Expose monopoles in lattice configurations in some gauge (usually in maximal abelian gauge)\cite{Suzuki}. Try to extract from them a monopole effective action in order to show that they undergo a Higgs 
    condensation below $T_c$ , but not above $T_c$.
    
    A short description  of the status of these investigations, in particular of the most recent progresses,
    will be the rest of my talk.  

    \section{THE DISORDER PARAMETER}
    I will shortly review the definition and the construction of the disorder parameter which detects 
  dual superconductivity of the vacuum \cite{DelDigPaf} \cite{DigPaff}. For the sake of simplicity I will
  do that for an U(1) gauge  theory in the language of the continuum formulation.
    
    The idea is to define an operator $\mu $ which carries non zero magnetic charge, and to focus on its 
 vacuum expectation value $\langle \mu \rangle$. A non zero $\langle \mu \rangle$ indicates that the magnetic charge of the vacuum is not defined, i. e. that there is condensation of monopoles, or Higgs breaking of  the magnetic gauge symmetry, which is nothing but dual superconductivity.  In a normal vacuum instead $\langle \mu \rangle = 0$ and the magnetic charge is superselected. 
 
 The definition of $\mu$ is :
 
 \begin{equation}
 \mu(\vec x, t)  =  exp[{ i q\over e^2} \int d^3 y \vec E(\vec y, t)\vec  b_{\perp}(\vec x - \vec y)]
 \end{equation}
 
 where $\vec b_{\perp}(\vec x) = {q\over 2} {{\vec x \wedge \vec n} \over {x(x-\vec x \vec n)}}$  is the vector potential describing the field produced by a monopole of charge  $q$ at a distance $\vec x$, in the transverse gauge ,
 
  $\nabla \vec b_{\perp} =0$  , $ \nabla \wedge \vec b_{\perp} = {q\over 2} {\vec x \over x^3} $+ Dirac  string along $\vec n$.
 
 Only the transverse part of $\vec E$ , $\vec E_{\perp}$ contributes to the convolution at the exponent of Eq(1) .$\vec E_{\perp}$ is the conjugate momentum to  $\vec A_{\perp}$ in whatever quantization procedure, so that  a formula analogous to the usual translation holds
 
 \begin{equation}
 exp(ipa)|x\rangle  = | x + a\rangle
 \end{equation}
 
 namely
 \begin{equation}
 \mu (\vec x, t) |\vec A_{\perp} (\vec z, t)\rangle = |\vec A_{\perp} (\vec z, t) + {1\over e}\vec b_{\perp}(\vec z -\vec x)\rangle
 \end{equation}
 Notice the factor  ${1\over e^2} $ at the exponent of Eq(1) : one  factor   ${1\over e} $ comes from the Dirac 
 quantization condition of the magnetic charge, the other one from the fact  the electric field as defined on a lattice  contains a multiplicative factor $e$ with respect to the conjugate momentum.
 
 The euclidean version of Eq(1) reads
 \begin{equation}
 \mu = exp(- \beta \Delta S)
 \end{equation}
 
 with $\beta = {1\over e^2}$  and     $\Delta S = \int d^3y \vec E(\vec y, t) \vec b_{\perp}(\vec x-\vec y)$
 
 The order parameter is finally\cite{DigPaff}
 \begin{equation}
 \langle \mu \rangle = { {\int [D A_{\mu}]  exp[-\beta (S + \Delta S)]}\over {\int [D A_{\mu}]  exp[-\beta S ]}}
 \end{equation}
 At $\beta =0$  $ \langle \mu \rangle =1$.
 
 It is convenient to define the susceptibility  $\rho $ 
 \begin{equation}
 \rho(\beta ) \equiv {\partial ln(\langle \mu \rangle)\over {\partial \beta}} = \langle S \rangle_{S}- \langle S + \Delta S\rangle_{S + \Delta S}
 \end{equation}
 where the brackets indicate average and the subscript the action used to define the weight.
 
 $\langle \mu \rangle$ can then be computed as
 \begin{equation}
 \langle \mu \rangle = exp(\int _{0} ^{\beta}\rho(\beta') d\beta')
 \end{equation}
 
  Compact $U(1)$ gauge theory with Wilson action in $3+1$ dimensions has a phase transition at a critical value $\beta_c \approx 1.01$ from a confined phase in which the Wilson loop obeys the area law , to a deconfined phase in which it obeys a perimeter law.  It is a theorem \cite{Frolich} that  $ \langle \mu \rangle \neq 0$ in the thermodynamical limit 
 in the confined phase and $ \langle \mu \rangle = 0$ in the deconfined one, which demonstrates that 
 for this system the mechanism of confinement is indeed dual superconductivity.
  This has been checked by numerical analysis via Montecarlo simulations \cite{DigPaff}.The result is shown in Figs (1) and (2).
  Fig (1) shows $\langle \mu \rangle$  versus $\beta$.
  \begin{figure}[ht]
\includegraphics[width=.5\textwidth,clip=]{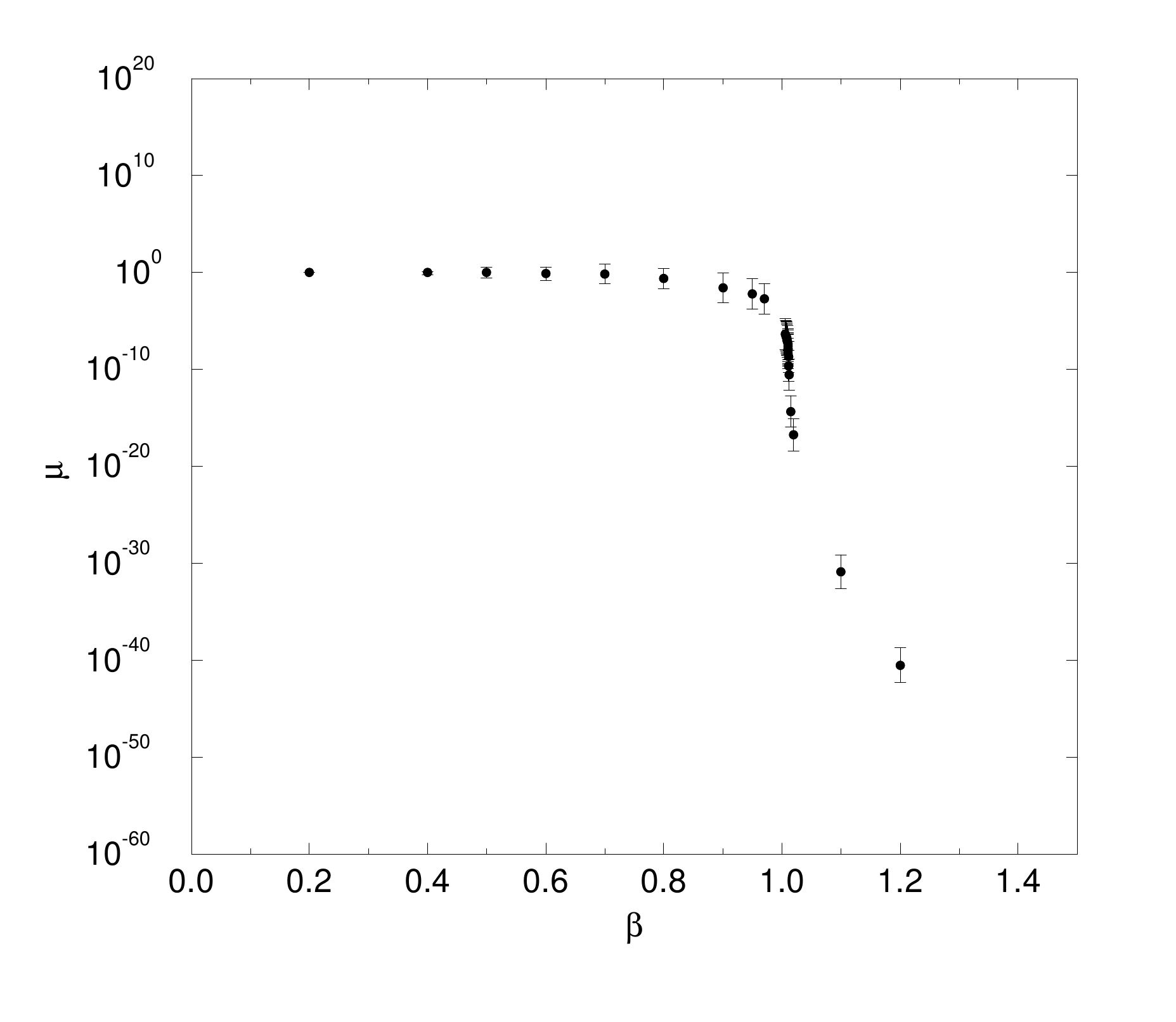}
 \caption{Fig 1 . $\langle \mu \rangle$ for U(1) lattice gauge theoryRef\cite {DigKyo}.
}
\end{figure}

Fig(2) shows the corresponding $\rho$ . It has a strong negative peak at the transition.

\begin{figure}
\includegraphics[height=.25\textheight]{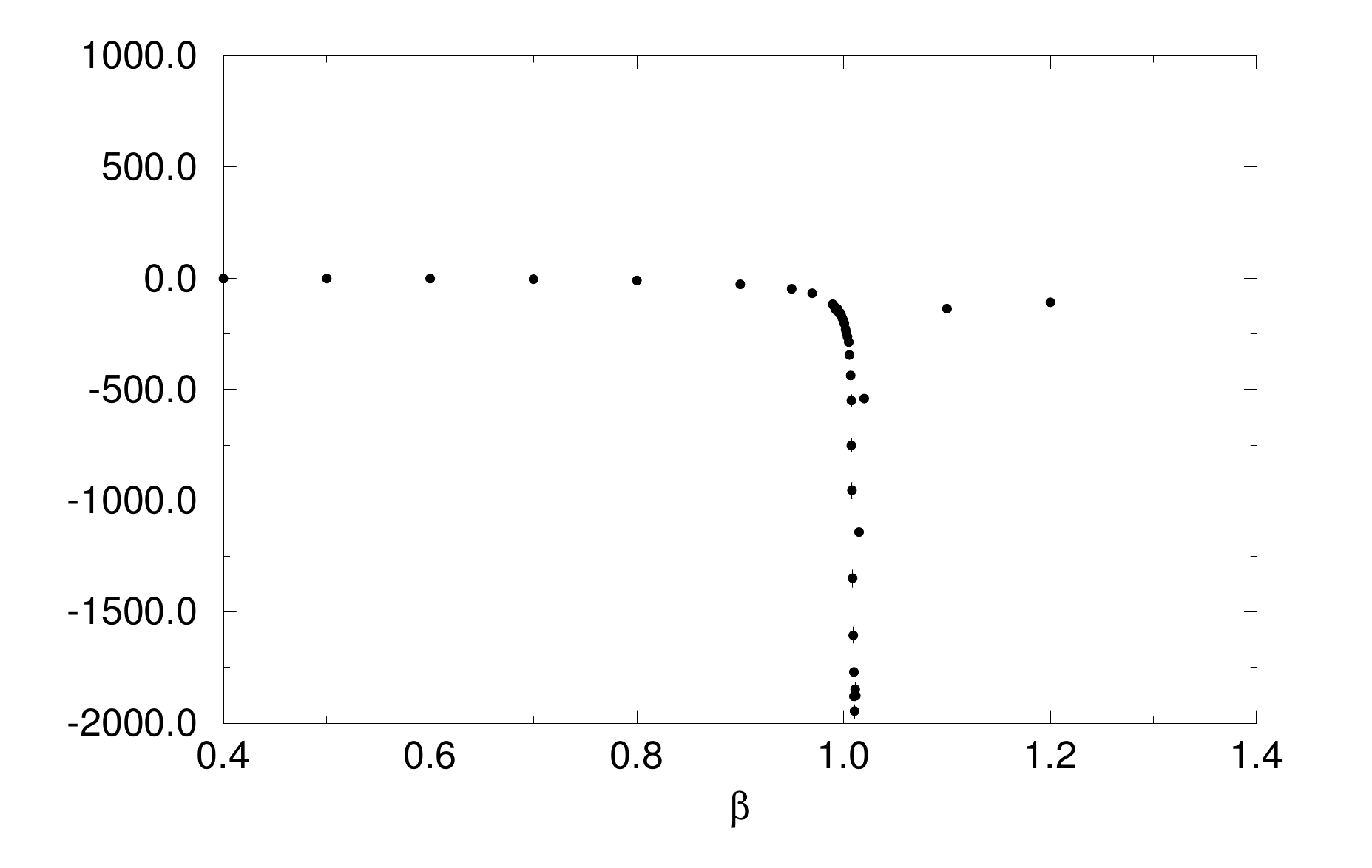}
\caption{Fig 2 . $\rho$ versus $\beta$ \cite{DigPaff}}
\end{figure}
The definition of the order parameter can be extended to the case of non-abelian gauge theories with \cite{CarDig2} and without quarks \cite{DigLuc1}\cite{DigLuc2}.
 For $SU(N)$ there are $(N - 1)$ different magnetic charges and order parameters $\langle \mu^a\rangle$ $(a = 1,..N-1)$. A priori they could depend on the abelian projection used to define the monopoles. However it has been shown both analytically\cite{Digp}\cite{DigPaff2} and  numerically\cite{CarDig1}
 that they are independent of it.
The behavior in the thermodynamical limit can be studied as follows:

1) For $\beta \leq \beta_c $ $\rho (\beta) $ tends to a finite limit as the volume $V\to \infty$.
    For large enough volumes it becomes volume independent. By use of Eq(7) it follows that
    $\langle \mu \rangle \neq 0$.
    
2) For $\beta \geq \beta_c $    $\rho  \approx -|c| L_s + c' $ with $L_s$ the linear size of the system. 
     As  $L_s \to \infty $, again by use of  Eq(7),  $\langle \mu \rangle \to 0$.
     
3) For   $\beta \approx \beta_c $  the correlation length $\lambda$ goes large as  $\lambda \approx \tau^   {-\nu}$  , with  $\tau \equiv (1- {T \over T_c})$ the reduced temperature and $\nu$ a critical index. The dependence on the lattice spacing becomes unimportant and there is scaling.

In formulae in the generic dependence 
\begin{equation}
\langle \mu \rangle = L_s^{\gamma} \Phi ({a\over \lambda},{L_s\over\lambda},mL_s^d)
\end{equation}
the ratio ${a\over \lambda}$ can be approximated by zero as $\beta \to \beta_c $, the variable 
${L_s \over \lambda}$ can be traded with the variable $\tau L_s^{1\over {\nu}}$ and therefore
\begin{equation}
\langle \mu \rangle = L_s^{\gamma} \Phi (0,\tau L_s^{1\over \nu},mL_s^d)
\end{equation}  In Eq's(8) and (9) $m$ is the quark mass.  If the theory is quenched (no quarks) the dependence on it disappears and by use of Eq(6) the scaling law follows for $\rho$
\begin{equation}
\rho/{L_s^{1\over \nu}} =f (\tau {L_s^{1\over \nu}})
\end{equation}
In the presence of quarks , keeping the first scaling variable $\tau L_s^{1\over \nu}$ fixed, at sufficiently large volumes the divergent factor in front  $L_s^{\gamma} $ must be compensated by the dependence 
on the second scaling variable and
\begin{equation} 
\langle \mu \rangle \approx m^{-\gamma\over d} \bar \Phi (\tau {L_s\over\lambda})
\end{equation}
whence by use of the definition of $\rho$ Eq(6) again the scaling law Eq (10) follows.

The scaling law Eq(10) (finite size scaling) allows to extract from numerical simulations the critical
index $\nu$, and with it information on the order and universality class of the phase transition.
 For pure $SU(2)$ and $SU(3)$  gauge theory the peak of $\rho$ coincides with that of the susceptibility of the Polyakov line and the value of $\nu$ is consistent with first order for $SU(3)$ and with second order for $SU(2)$
 in the universality class of the Ising in 3d, again in agreement with the determination from the Polyakov line\cite{DigLuc1}\cite{DigLuc2}.
 
 I will now illustrate with some figures the arguments discussed above. 
 Fig(3) shows the behavior of $\rho$ below $\beta_c$ for QCD with two flavors. The independence 
 on the volume is clearly seen, implying a finite thermodynamical  limit of $\rho$ or a non zero value of 
$\langle \mu \rangle$, i.e. dual superconductivity below $\beta_c$.
 \begin{figure}
\includegraphics[height=.25\textheight]{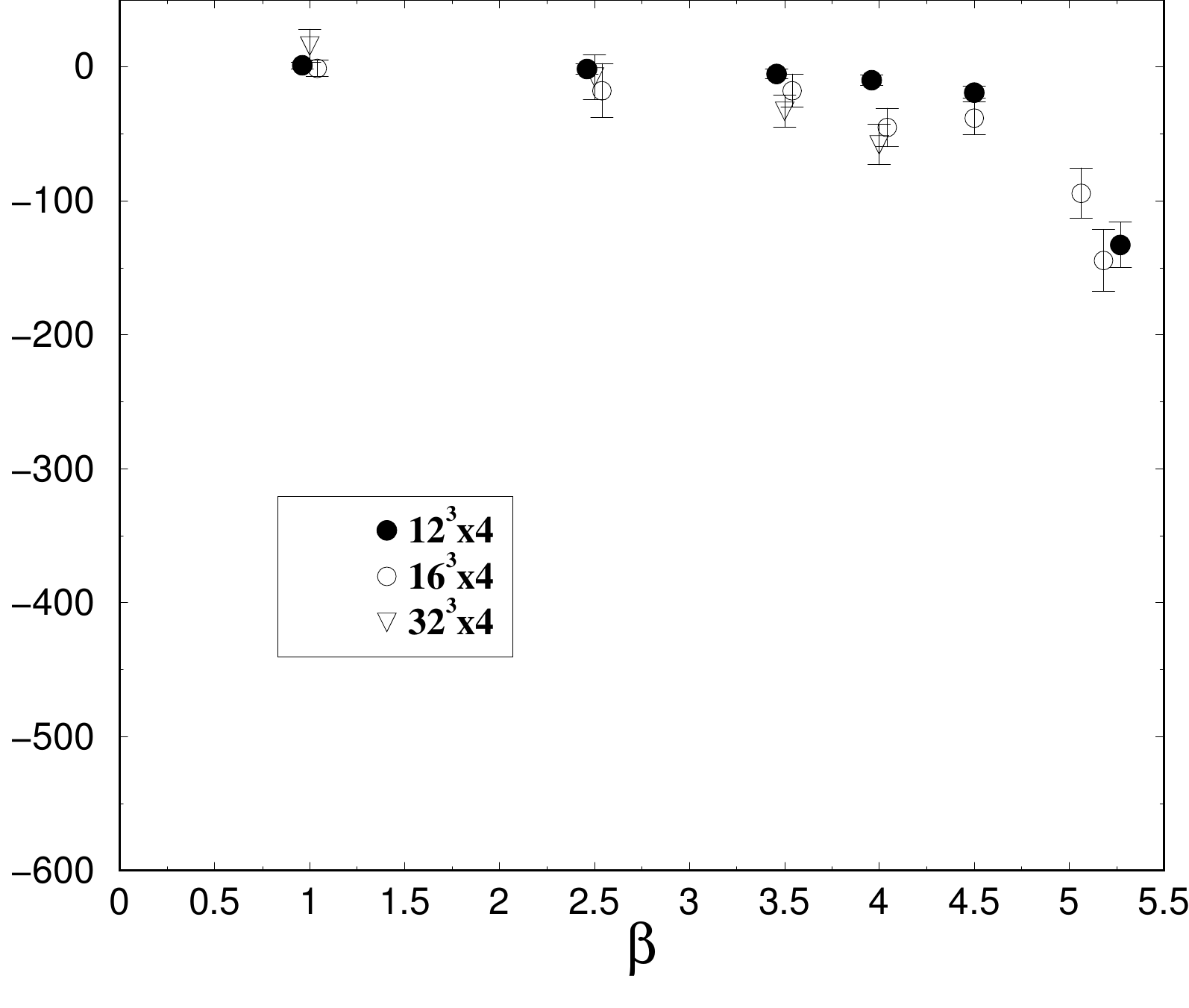}
 \caption{Fig. 3 Strong coupling behavior of $\rho$ at various lattice sizes and $am=0.1335$\cite{CarDig2}}
\end{figure}

Fig(4) shows the volume dependence of $\rho$ in the deconfined phase \cite{DelDigLuc}. Whenever the magnetic charge carried by $\mu$ is non zero $\rho$ diverges linearly to $-\infty$ with the size $L_s$, implying that $\mu$ is strictly zero in the thermodynamical limit, or that magnetic $U(1)$ is a Wigner symmetry, and superconductivity has disappeared.

 \begin{figure}
\includegraphics[height=.25\textheight]{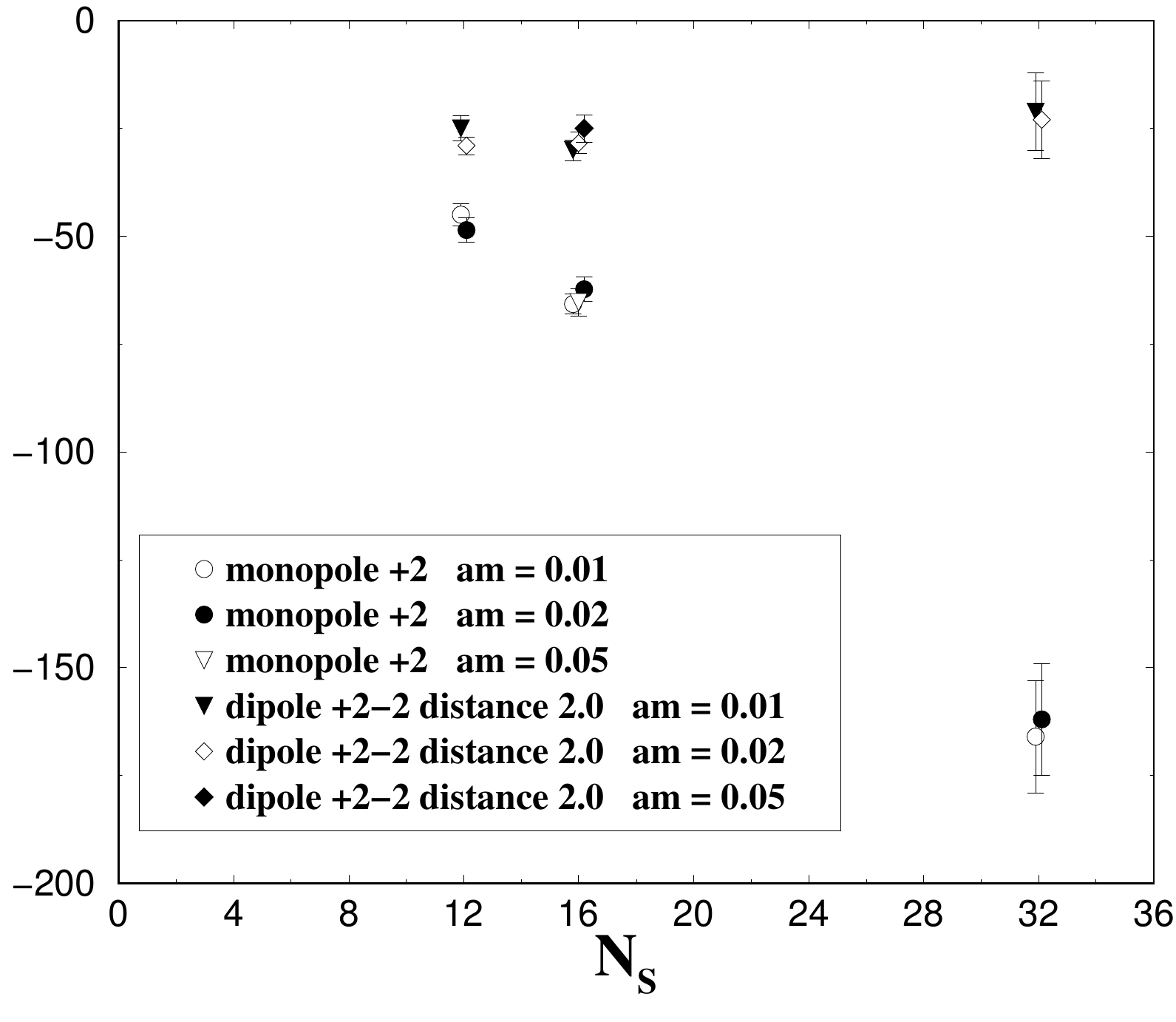}
 \caption{Fig. 4 Volume dependence of $\rho$ in the deconfined phase for different values of the magnetic charge.\cite{DelDigLuc}}
\end{figure}

Fig(5) shows the negative peak of $\rho$  at the critical point, and the chiral condensate 
superimposed. The peak is located at the temperature at which chiral symmetry is restored.
Chiral and deconfining transitions coincide \cite{CarDig2}.

\begin{figure}
\includegraphics[height=.25\textheight]{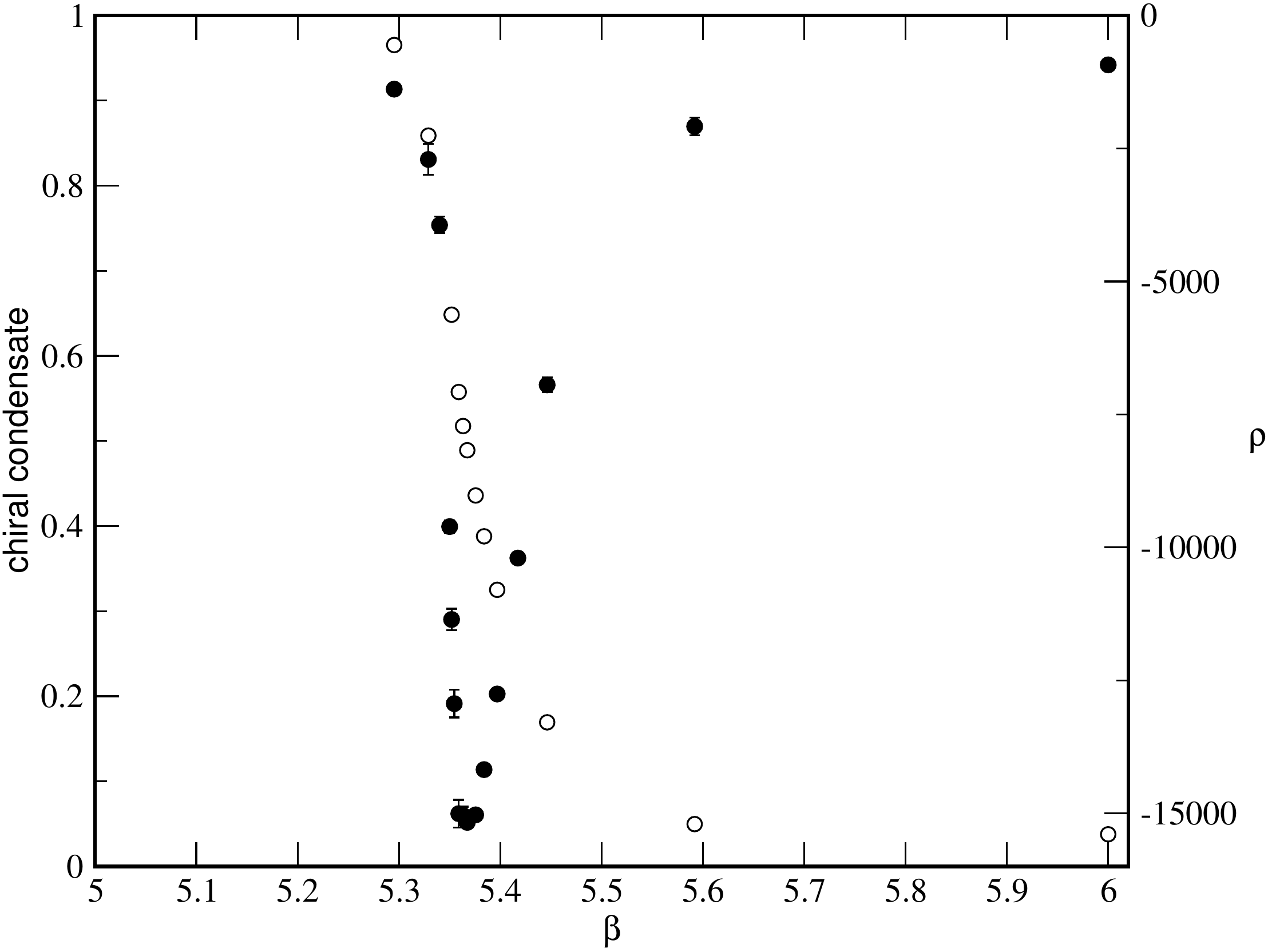}
 \caption{Fig. 5 $\rho$ (rigth vertical scale, full dots) and $\langle \bar \psi \psi \rangle$ (left vertical scale, open dots) at the critical point.  $N_f=2 $ $am=.1335$\cite{CarDig2}}
\end{figure}

Fig(6) shows the scaling of $\rho$ in the critical region for  $\nu = {1\over 3} $ , which corresponds to first order.  Data at different  $L_s$  are rescaled as in Eq(10), and the different curves fall on each other,
indicating that the transition is first order. There is practically no dependence on the quark mass , in agreement with Eq(11).

\begin{figure}
\includegraphics[height=.25\textheight]{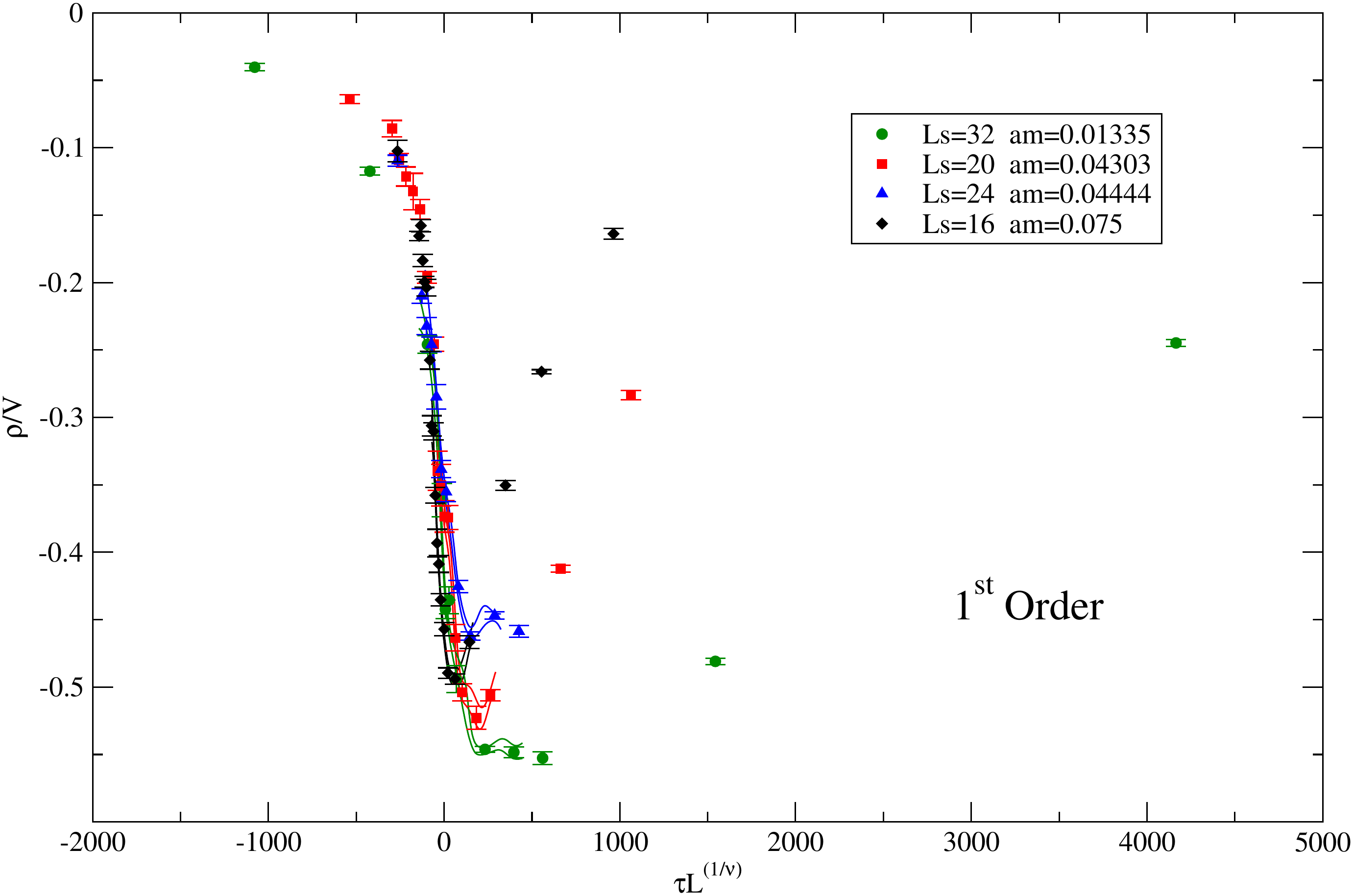}
 \caption{Fig. 6. Scaling of  $\rho$ assuming first order for the deconfining transition.) at the critical point.  $N_f=2 $ \cite{DelDigPic1}}
\end{figure}

The scaling Eq(10) with the indexes of $O(4)$ universality class is shown in Fig(7) . The curves should overlap if the $O(4)$ universality class were correct, and they do not.

\begin{figure}
\includegraphics[height=.25\textheight]{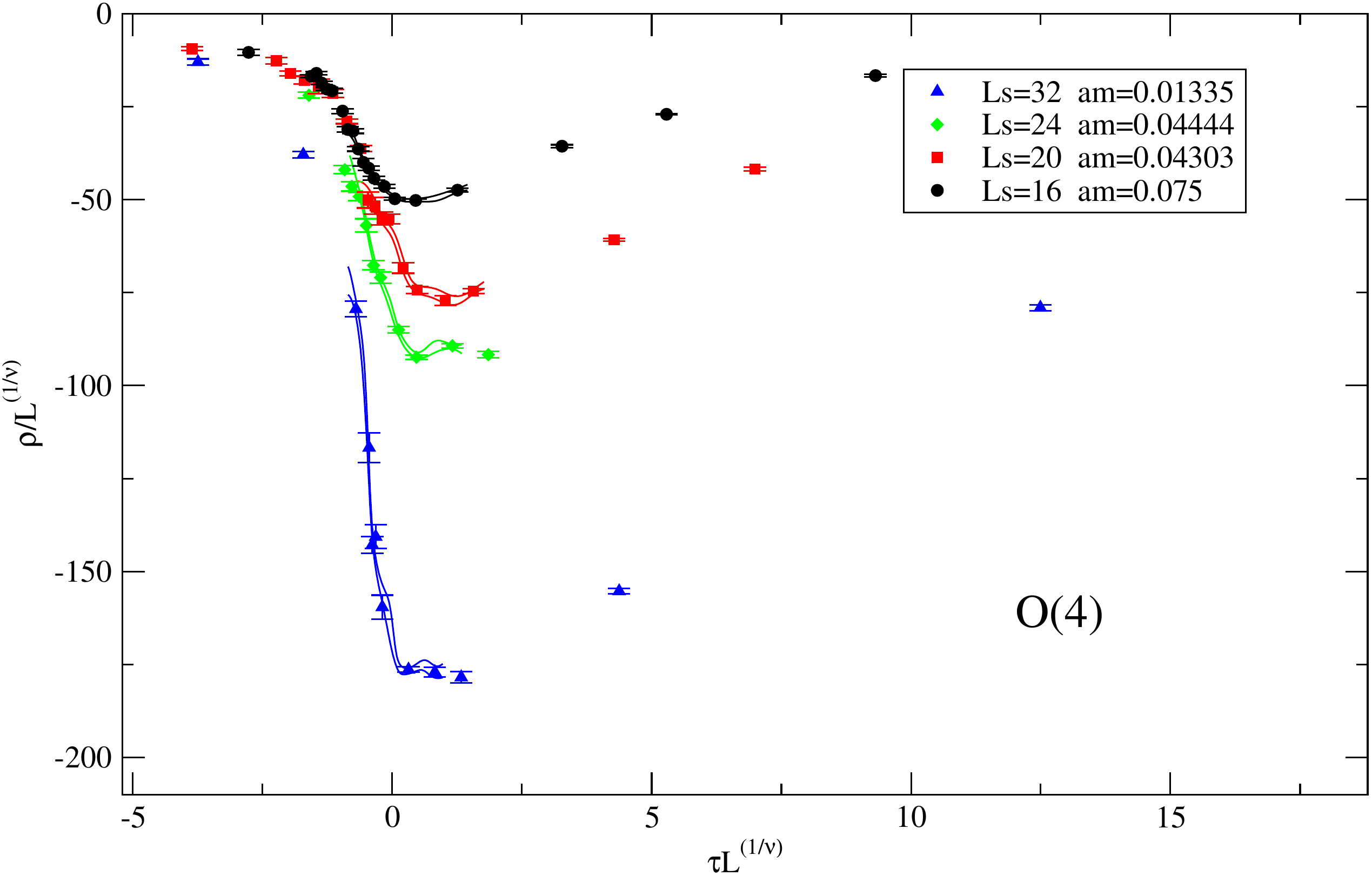}
 \caption{Fig. 7. Scaling of  $\rho$ at the critical point for  $N_f=2 $ assuming second order $O(4)$ for the chiral transition.  \cite{DelDigPic1}}
\end{figure}

This result is in agreement with a systematic and careful finite size scaling analysis of the specific heat 
and of the chiral susceptibility done on the same system \cite{DelDigPic1}\cite{DelDigPic2}.
Assuming that the order parameters for the chiral transition are the $vev$ of scalar and pseudoscalar fields the chiral transition for $N_f$ $QCD$ is expected either to be first order at $m=0$ and then the transition is also first order at small, non zero values of the mass, or to be second order in the universality class of $O(4)$ , and then it is a crossover at $m\neq 0$ \cite{PisWil}. This is a fundamental issue: if the transition is first order deconfinement can be an order-disorder transition , i.e. a change of symmetry, and the choice of Nature is the natural one, in the sense explained in Sect . 1; if instead it is second order $O(4)$   it is a crossover at $m\neq 0$ and then one has to explain the origin of the inhibition factor $\leq 10^{-15}$ across a continuous transformation.

\section{MONOPOLE DOMINANCE	}
For many years there has been a common belief in the community of lattice theorists
 that not all the abelian projections were on the same footing: the maximal abelian projection
 for some reason was privileged, since apparently only in this projection the dominance of the abelian degrees of freedom, and specifically of the monopoles  was realized\cite{SSuz}. 
 
 It was instead clear in the approach based on symmetry that all abelian projections are equivalent 
 \cite{Digp}\cite{DigPaff2}\cite{CarDig1}.
 Recently an important result has been  obtained \cite{Suz} by improved numerical techniques:
 Monopole dominance and abelian dominance hold in all abelian projections, and also without fixing the gauge.This was already noticed in \cite{CarDig1} and \cite{CeaCos}. 

\subsection{CONCLUSIONS}
One of the basic questions in confinement is the determination of the order of the deconfining transition in QCD, and it should be studied with great care. Systems like QCD with quarks in the adjoint representation can provide important insight : the deconfining and chiral transition occur in this case at different temperatures \cite {Kar}and are first order and continuous respectively \cite{cosdel} . Deconfinement is again well described by the order parameter $\langle \mu \rangle$.

Dual superconductivity as the mechanism of color confinement receives increasing evidence from
lattice simulations: if confirmed it would support the natural choice based on symmetry.


\begin{theacknowledgments}
I thank  G.Cossu, M. D'Elia, B. Lucini, G. Paffuti, C. Pica for collaboration.
  \end{theacknowledgments}

\end{document}